\documentclass[manyauthors,nocleardouble,COMPASS]{cernphprep}

\usepackage{bm}
\usepackage{color}

\usepackage{amssymb}
\usepackage{amsmath}
\usepackage{amsbsy}
\usepackage{times}
\usepackage{epsfig}
\usepackage{colordvi}
\usepackage{graphicx}
\usepackage{wrapfig,rotating}
\usepackage[numbers, square, comma, sort&compress]{natbib}
\usepackage{hyperref} 
\usepackage{multicol}
\usepackage{booktabs}
\usepackage{multirow}
\usepackage{lineno}
\RequirePackage[T1]{fontenc}
 
\makeatletter

\DeclareSymbolFont{letters}     {OML}{cmm}{m}{it}
\DeclareSymbolFont{symbols}     {OMS}{cmsy}{m}{n}
\DeclareSymbolFont{largesymbols}{OMX}{cmex}{m}{n}
\newcommand{\Jpsi}{J\!/\!\psi}

\newcommand{\GeV}{\mathrm{GeV}}

\begin{document}

\begin{titlepage}

\PHnumber{2014--180}
\PHdate{July 21, 2014}

\title{Search for exclusive photoproduction of Z$_c^{\pm}$(3900) at COMPASS}

\Collaboration{The COMPASS Collaboration}
\ShortAuthor{The COMPASS Collaboration}

\begin{abstract}
A search for the exclusive production of the $Z_c^{\pm}(3900)$ hadron by virtual
photons has been performed in the channel $Z_c^{\pm}(3900)\rightarrow J/\psi
\pi^{\pm}$.  The data cover the range from 7~GeV to 19~GeV in the centre-of-mass
energy of the photon-nucleon system. The full set of the COMPASS data set
collected with a muon beam between 2002 and 2011 has been used.  An upper limit
for the ratio $BR(Z_c^{\pm}(3900)\rightarrow J/\psi \pi^{\pm} )\times \sigma_{
  \gamma~N \rightarrow Z_c^{\pm}(3900)~ N} /\sigma_{ \gamma~N \rightarrow
  J/\psi~ N}$ of $3.7\times10^{-3}$ has been established at the confidence level
of 90\%.


\end{abstract}
\vfill
\Submitted{(to be submitted to Phys. Lett. B)}

\end{titlepage}

{\pagestyle{empty}
%
%

\section*{The COMPASS Collaboration}
\label{app:collab}
\renewcommand\labelenumi{\textsuperscript{\theenumi}~}
\renewcommand\theenumi{\arabic{enumi}}
\begin{flushleft}
C.~Adolph\Irefn{erlangen},
R.~Akhunzyanov\Irefn{dubna}, 
M.G.~Alexeev\Irefn{turin_u},
G.D.~Alexeev\Irefn{dubna}, 
A.~Amoroso\Irefnn{turin_u}{turin_i},
V.~Andrieux\Irefn{saclay},
V.~Anosov\Irefn{dubna}, 
A.~Austregesilo\Irefnn{cern}{munichtu},
B.~Bade{\l}ek\Irefn{warsawu},
F.~Balestra\Irefnn{turin_u}{turin_i}
% ...
\end{flushleft}

%
%

\begin{Authlist}
\item \Idef{bielefeld}{Universit\"at Bielefeld, Fakult\"at f\"ur Physik, 33501 Bielefeld, Germany\Arefs{f}}
\item \Idef{bochum}{Universit\"at Bochum, Institut f\"ur Experimentalphysik, 44780 Bochum, Germany\Arefs{f}\Arefs{ll}}
\item \Idef{bonniskp}{Universit\"at Bonn, Helmholtz-Institut f\"ur  Strahlen- und Kernphysik, 53115 Bonn, Germany\Arefs{f}}
\item \Idef{bonnpi}{Universit\"at Bonn, Physikalisches Institut, 53115 Bonn, Germany\Arefs{f}}
\item \Idef{brno}{Institute of Scientific Instruments, AS CR, 61264 Brno, Czech Republic\Arefs{g}}
\item \Idef{calcutta}{Matrivani Institute of Experimental Research \& Education, Calcutta-700 030, India\Arefs{h}}
\item \Idef{dubna}{Joint Institute for Nuclear Research, 141980 Dubna, Moscow region, Russia\Arefs{i}}
\item \Idef{erlangen}{Universit\"at Erlangen--N\"urnberg, Physikalisches Institut, 91054 Erlangen, Germany\Arefs{f}}
\item \Idef{freiburg}{Universit\"at Freiburg, Physikalisches Institut, 79104 Freiburg, Germany\Arefs{f}\Arefs{ll}}
\item \Idef{cern}{CERN, 1211 Geneva 23, Switzerland}
\item \Idef{liberec}{Technical University in Liberec, 46117 Liberec, Czech Republic\Arefs{g}}
\item \Idef{lisbon}{LIP, 1000-149 Lisbon, Portugal\Arefs{j}}
\item \Idef{mainz}{Universit\"at Mainz, Institut f\"ur Kernphysik, 55099 Mainz, Germany\Arefs{f}}
\item \Idef{miyazaki}{University of Miyazaki, Miyazaki 889-2192, Japan\Arefs{k}}
\item \Idef{moscowlpi}{Lebedev Physical Institute, 119991 Moscow, Russia}
\item \Idef{munichlmu}{Ludwig-Maximilians-Universit\"at M\"unchen, Department f\"ur Physik, 80799 Munich, Germany\Arefs{f}\Arefs{l}}
\item \Idef{munichtu}{Technische Universit\"at M\"unchen, Physik Department, 85748 Garching, Germany\Arefs{f}\Arefs{l}}
\item \Idef{nagoya}{Nagoya University, 464 Nagoya, Japan\Arefs{k}}
\item \Idef{praguecu}{Charles University in Prague, Faculty of Mathematics and Physics, 18000 Prague, Czech Republic\Arefs{g}}
\item \Idef{praguectu}{Czech Technical University in Prague, 16636 Prague, Czech Republic\Arefs{g}}
\item \Idef{protvino}{State Scientific Center Institute for High Energy Physics of National Research Center `Kurchatov Institute', 142281 Protvino, Russia}
\item \Idef{saclay}{CEA IRFU/SPhN Saclay, 91191 Gif-sur-Yvette, France\Arefs{ll}}
\item \Idef{telaviv}{Tel Aviv University, School of Physics and Astronomy, 69978 Tel Aviv, Israel\Arefs{m}}
\item \Idef{triest_u}{University of Trieste, Department of Physics, 34127 Trieste, Italy}
\item \Idef{triest_i}{Trieste Section of INFN, 34127 Trieste, Italy}
\item \Idef{triest_ictp}{Abdus Salam ICTP, 34151 Trieste, Italy}
\item \Idef{turin_u}{University of Turin, Department of Physics, 10125 Turin, Italy}
\item \Idef{turin_p}{University of Eastern Piedmont, 15100 Alessandria, Italy}
\item \Idef{turin_i}{Torino Section of INFN, 10125 Turin, Italy}
\item \Idef{warsaw}{National Centre for Nuclear Research, 00-681 Warsaw, Poland\Arefs{n} }
\item \Idef{warsawu}{University of Warsaw, Faculty of Physics, 00-681 Warsaw, Poland\Arefs{n} }
\item \Idef{warsawtu}{Warsaw University of Technology, Institute of Radioelectronics, 00-665 Warsaw, Poland\Arefs{n} }
\item \Idef{yamagata}{Yamagata University, Yamagata, 992-8510 Japan\Arefs{k} }
\end{Authlist}
%
%
\vspace*{-\baselineskip}\renewcommand\theenumi{\alph{enumi}}
\begin{Authlist}
\item \Adef{a}{Also at Instituto Superior T\'ecnico, Universidade de Lisboa, Lisbon, Portugal}
\item \Adef{aa}{Also at Department of Physics, Pusan National University, Busan 609-735, Republic of Korea and at Physics Department, Brookhaven National Laboratory, Upton, NY 11973, U.S.A. }
\item \Adef{bb}{Supported by the DFG Research Training Group Programme 1102  ``Physics at Hadron Accelerators''}
\item \Adef{b}{Also at Chubu University, Kasugai, Aichi, 487-8501 Japan\Arefs{k}}
\item \Adef{c}{Also at KEK, 1-1 Oho, Tsukuba, Ibaraki, 305-0801 Japan}
\item \Adef{c1c}{Present address: Universit\"at Bonn, Helmholtz-Institut f\"ur Strahlen- und Kernphysik, 53115 Bonn, Germany}
\item \Adef{cc}{Also at Moscow Institute of Physics and Technology, Moscow Region, 141700, Russia}
\item \Adef{x}{present address: RWTH Aachen University, III. Physikalisches Institut, 52056 Aachen, Germany}
\item \Adef{f}{Supported by the German Bundesministerium f\"ur Bildung und Forschung}
\item \Adef{g}{Supported by Czech Republic MEYS Grants ME492 and LA242}
\item \Adef{h}{Supported by SAIL (CSR), Govt.\ of India}
\item \Adef{i}{Supported by CERN-RFBR Grants 08-02-91009 and 12-02-91500}
\item \Adef{j}{\raggedright Supported by the Portuguese FCT - Funda\c{c}\~{a}o para a Ci\^{e}ncia e Tecnologia, COMPETE and QREN, Grants CERN/FP/109323/2009, CERN/FP/116376/2010 and CERN/FP/123600/2011}
\item \Adef{k}{Supported by the MEXT and the JSPS under the Grants No.18002006, No.20540299 and No.18540281; Daiko Foundation and Yamada Foundation}
\item \Adef{l}{Supported by the DFG cluster of excellence `Origin and Structure of the Universe' (www.universe-cluster.de)}
\item \Adef{ll}{Supported by EU FP7 (HadronPhysics3, Grant Agreement number 283286)}
\item \Adef{m}{Supported by the Israel Science Foundation, founded by the Israel Academy of Sciences and Humanities}
\item \Adef{n}{Supported by the Polish NCN Grant DEC-2011/01/M/ST2/02350}
\item [{\makebox[2mm][l]{\textsuperscript{*}}}] Deceased
\end{Authlist}

\newpage

The $Z_c^{\pm}(3900)$ state was recently discovered by the BES-III and Belle
Collaborations in $e^+e^-\rightarrow \pi^+\pi^- \Jpsi$ reactions at
$\sqrt{s}=4.26~\GeV$ \cite{BES3, Belle} via the decay channel
\begin{equation}
\label{decay}
 Z_c^{\pm}(3900)\rightarrow \Jpsi \pi^{\pm}.
\end{equation}
It has been interpreted as a tetraquark state \cite{tet1,tet2,tet3,tet4},
although other explanations like a molecular state
\cite{mol1,mol2,mol3,mol4,mol5}, a cusp effect \cite{cusp} and an
initial-single-pion-emission mechanism \cite{piem} were also proposed. According
to the vector meson dominance (VMD) model, a photon may behave like a $\Jpsi$ so
that a $Z_c^{\pm}(3900)$ can be produced by the interaction of an incoming
photon with a virtual charged pion provided by the target nucleon
\begin{equation}
\label{reaction}
\gamma~N \rightarrow Z_c^{\pm}(3900)~N.
\end{equation}
The corresponding diagram is shown in Fig.~\ref{fig:diag}a.

Based on the VMD model, the authors of Ref.~\cite{MAIN} predict a sizable cross
section of the reaction in Eq.~(\ref{reaction}) for $\sqrt{s_{\gamma N}}\sim
10~\GeV$. Under the assumption that the decay channel of Eq.~(\ref{decay}) is
dominant and that the total width $\Gamma_{\mbox{tot}}$ of the $Z_c^{\pm}(3900)$
particle is 46~MeV/$c^2$, as measured by BES-III, the cross section reaches a
maximum value of 50~nb to 100~nb at $\sqrt{s_{\gamma N}}= 7~\GeV$. The $\Jpsi$
production in photon-nucleon interactions at COMPASS covers the range
$\sqrt{s_{\gamma N}}$ from 7 GeV to 19 GeV and thus can be used to also study
$Z_c^{\pm}(3900)$ production and to estimate the partial width $\Gamma_{\Jpsi
  \pi}$ of the decay channel $Z_c^{\pm}(3900)\rightarrow \Jpsi \pi^{\pm}$.

\begin{figure}[h]
 \begin{center}
  \includegraphics[width=220px]{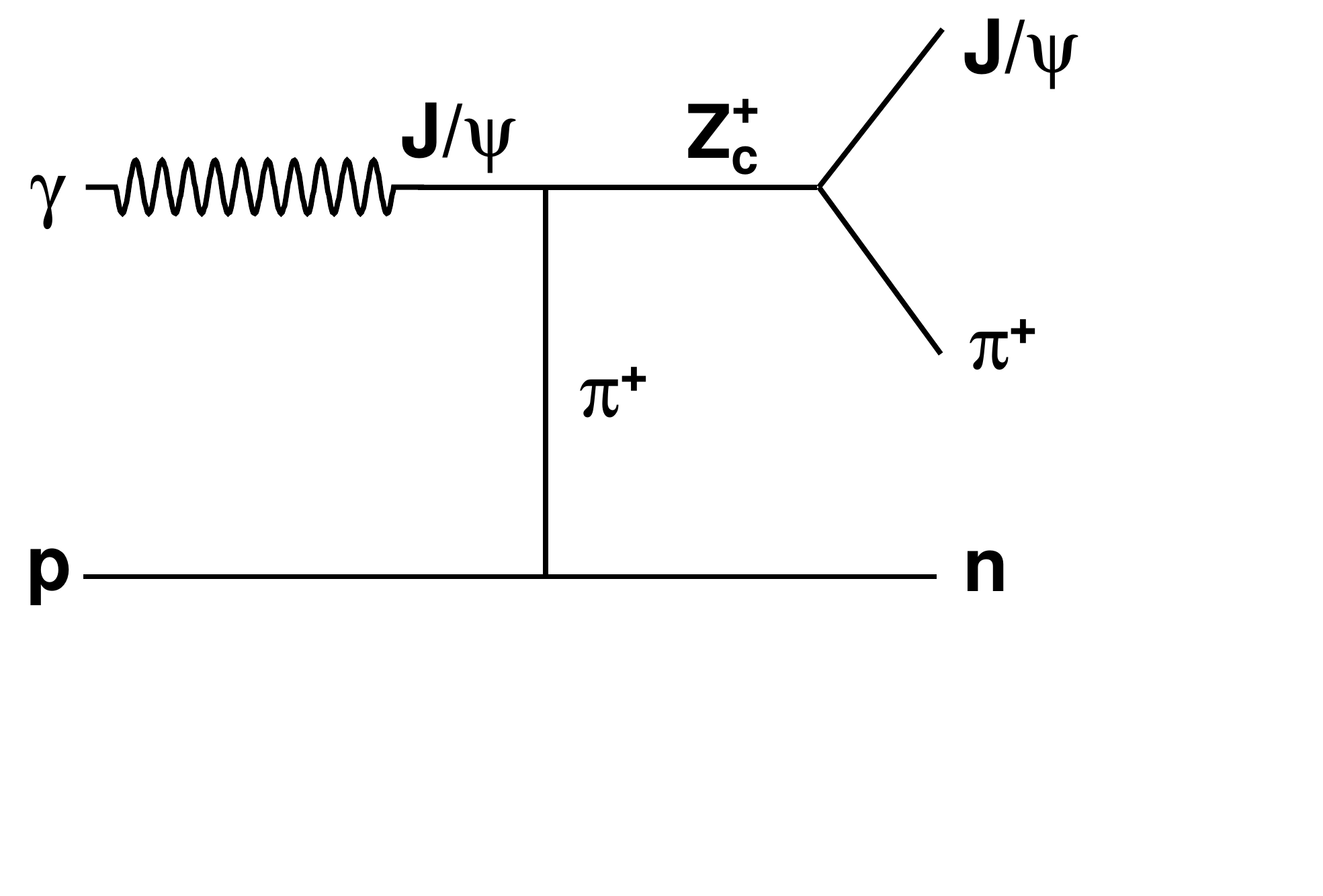}
  \includegraphics[width=220px]{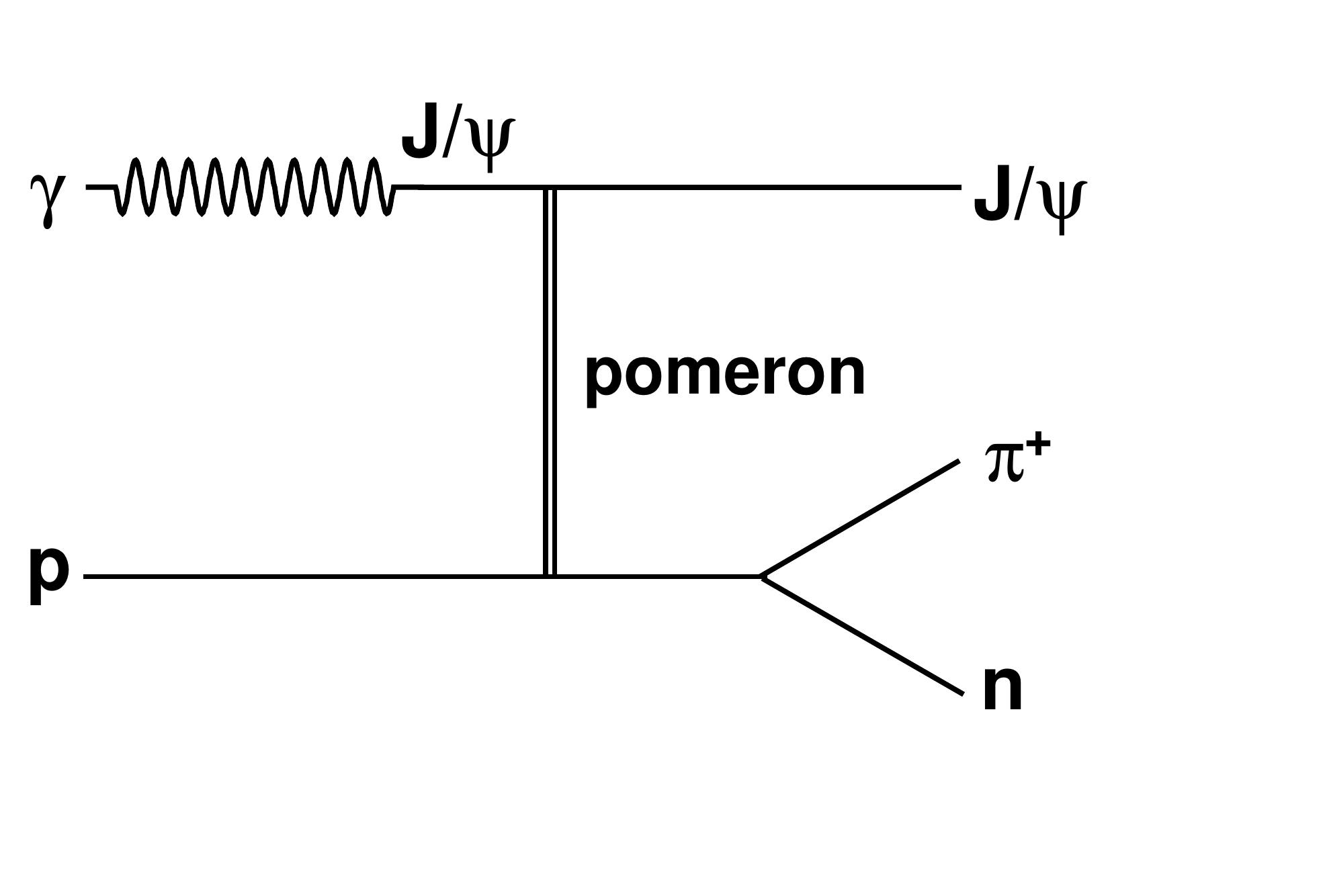}\\
  $\quad$(a)\hspace{0.4\textwidth}(b)$\quad$     \\      
   \end{center}
  \caption{\label{fig:diag}
Diagrams for (a) $Z_c^+(3900)$ production via virtual $\pi^+$ exchange and (b)
$\Jpsi \pi^{+}$ production via pomeron exchange.  }
\end{figure}

The COMPASS experiment \cite{Abbon:2007pq} is situated at the M2 beam line of
the CERN Super Proton Synchrotron.  The data used in the present analysis were
obtained scattering positive muons of 160~$\GeV/c$ (2002-2010) or 200~$\GeV/c$
momentum (2011) off solid $^6$LiD (2002-2004) or NH$_3$ targets (2006-2011).
The longitudinally or transversely polarised targets consisted of two
(2002--2004) or three (2006--2011) cylindrical cells placed along the beam
direction. Polarisation effects were canceled out by combining data with
opposite polarisation orientations. Particle tracking and identification were
performed in a two-stage spectrometer, covering a wide kinematic range.  The
trigger system comprises hodoscope counters and hadron calorimeters. Beam halo
was rejected by veto counters upstream of the target.

In the analysis presented in this Letter, the reaction
\begin{equation}
\label{reaction4}
\mu^+~N \rightarrow \mu^+ Z_c^{\pm}(3900)~N \rightarrow \mu^+\Jpsi \pi^{\pm}
N\rightarrow \mu^+\mu^+\mu^- \pi^{\pm} N
\end{equation}
was searched for.  In order to select samples of exclusive $\mu^+ \Jpsi
\pi^{\pm}$ events, a reconstructed vertex in the target region with an incoming
beam track and three outgoing muon tracks (two positive and one negative) is
required.  Tracks are attributed to muons if they cross more than 15 radiation
lengths of material.  Only the events with exactly three muons and one pion in
the final state were selected.  A pair of muons is treated as $\Jpsi$ candidate
if the difference between its reconstructed mass $M_{\mu^+\mu^-}$
(Fig.~\ref{fig:kin}a) and the nominal $\Jpsi$ mass is less than 150~MeV/$c^2$.
In case both $\mu^+\mu^-$ combinations satisfy this condition, the event is
rejected.  Except for the tiny recoil of the target nucleon, the sum of the
scattered muon energy, $E_{\mu'}$, and the energies of produced $\Jpsi$ and
$\pi^{\pm}$ mesons, $E_{\Jpsi}$ and $E_{\pi^{\pm}}$, should be equal to the beam
energy $E_b$ for the exclusive reaction of Eq.~(\ref{reaction4}). The
distribution of events as a function of the energy balance $\Delta
E=E_{\mu'}+E_{\Jpsi}+E_{\pi^{\pm}}-E_{b}$ is presented in Fig.~\ref{fig:kin}b.
With the experimental energy resolution of about 3~GeV, the energy balance is
required to be $|\Delta E|<$10~GeV.  The distribution of the negative squared
four-momentum transfer $Q^2= -(P_b-P_{\mu'})^2$ is shown in
Fig.~\ref{fig:kin2}a.  Here $P_{\mu'}$ and $P_{b}$ are four-momenta of the
scattered and incident muons, respectively.  The momentum of the produced pion
is required to be larger than 2~$\GeV/c$ in order to reduce the background of
exclusive events with a $\Jpsi$ and a $\pi^{\pm}$ in the final state produced
via pomeron exchange (Fig.~\ref{fig:diag}b). The total number of selected
$\mu^+\Jpsi \pi^+$ and $\mu^+\Jpsi \pi^-$ events is 565 and 405,
respectively. The distribution of the centre-of-mass energy of the
photon-nucleon system $\sqrt{s_{\gamma N}}$ is shown in Fig.~\ref{fig:kin2}b.

\begin{figure}
 \begin{center}
  \includegraphics[width=220px]{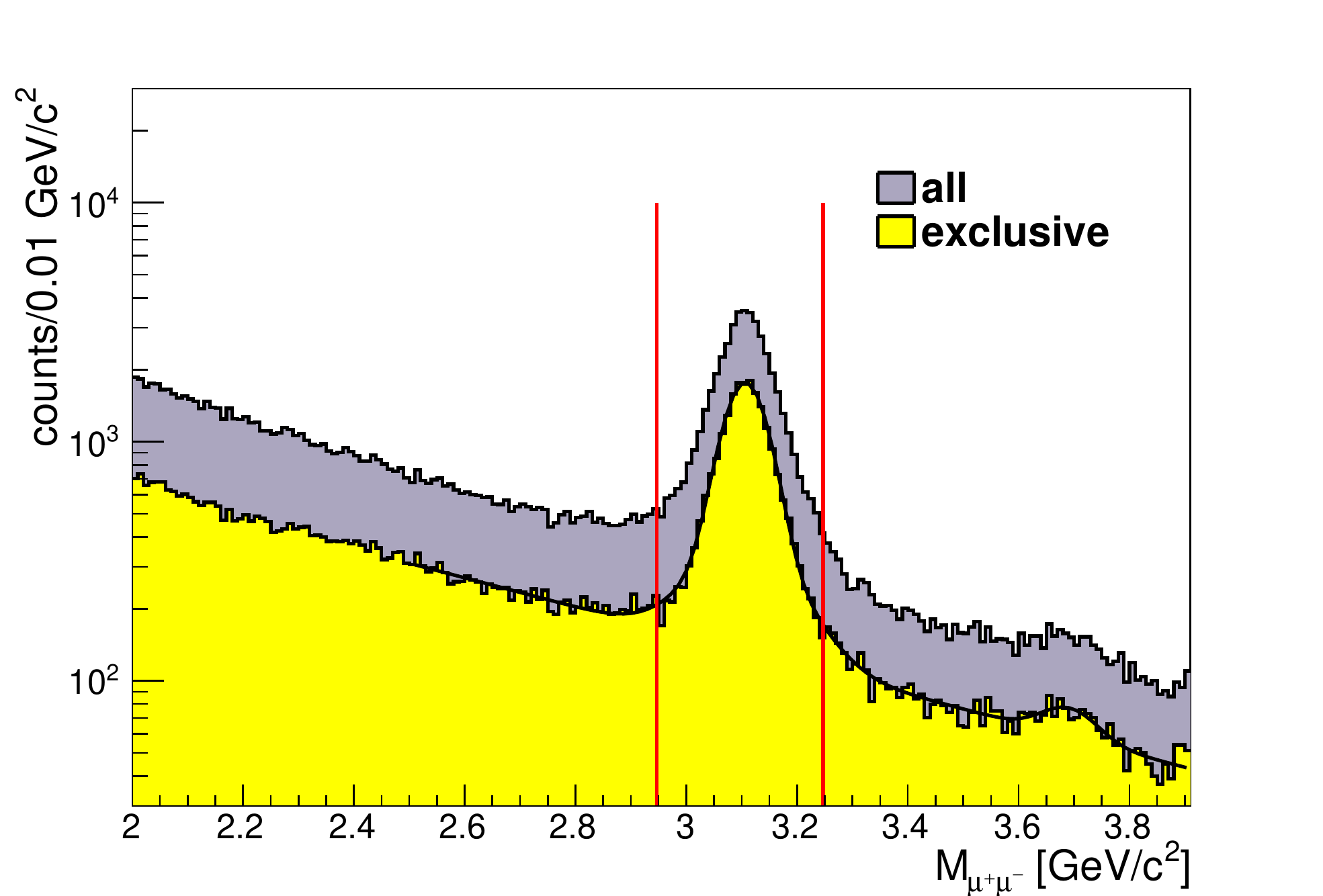}
  \includegraphics[width=220px]{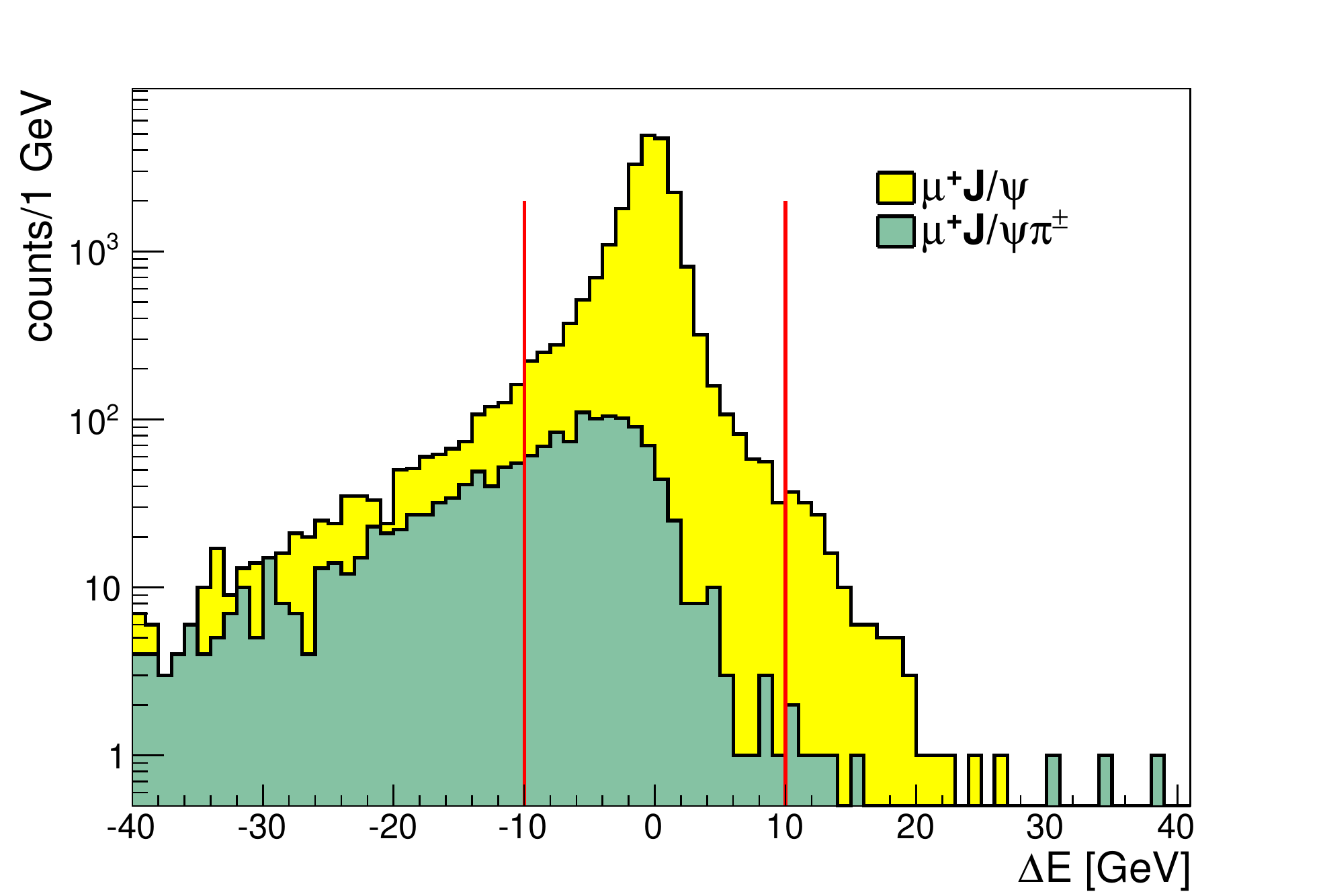}\\
  (a)\hspace{0.4\textwidth}(b)$\quad$ 
   \end{center}
  \caption{\label{fig:kin}
(a) The dimuon mass distribution for all dimuons produced in muon-nucleon
    scattering (blue, upper curve), and for exclusively produced dimuons
    (yellow, lower curve). (b) Distribution for the energy balance $\Delta E$ in
    the reactions Eq.~(7) (yellow, upper curve) and Eq.~(3) (green, lower
    curve).  }
\end{figure}

The mass spectrum for $\Jpsi \pi^{\pm}$ events is shown in
Fig.~\ref{fig:kin3}a. It does not exhibit any statistically significant resonant
structure around 3.9~GeV/$c^2$.  The observed number of events $N_{\Jpsi\pi}$ in
the signal range $3.84~\GeV/c^2<M_{\Jpsi \pi^+}<3.96~\GeV/c^2$ is treated as
consisting of an a priori unknown $Z_c^{\pm}(3900)$ signal $N_{Z_c}$ and a
background contribution $N_{bkg}$. According to the method described in
Ref.~\cite{peak}, the probability density function $g(N_{Z_c})$ is given by
\begin{equation}
g(N_{Z_c})=n
\int_{0}^{\infty}\frac{e^{-(N_{Z_c}+N_{bkg})}(N_{Z_c}+N_{bkg})^{N_{Z_c}}}{N_{Z_c}!}f(N_{bkg})dN_{bkg},
\end{equation}
where $n$ is a normalization constant and the probability density function
$f(N_{bkg})$, assumed to be Gaussian, describes the background contribution in
the signal interval. The mean value and the Gaussian width of $f(N_{bkg})$ are
estimated by fitting a sum of two exponential functions ($A\cdot e^{-a M_{\Jpsi
    \pi}} + B\cdot e^{-b M_{\Jpsi \pi}}$) to the $\Jpsi \pi^{\pm}$ mass spectrum
in the range $3.3~\GeV/c^2<M_{\Jpsi \pi^+}< 6.0~\GeV/c^2$ excluding the signal
region. The fitted function is shown as a line in Fig.~\ref{fig:kin3}a. The
number of expected background events in the signal region is $49.7\pm3.4$ while
51 is observed.  The upper limit $N^{UL}_{Z_c}$ for the number of produced
$Z^{\pm}_c(3900)$ events corresponding to a confidence level of $CL=90\%$ is
then determined from the expression
\begin{equation}
\int_0^{N^{UL}_{Z_c}} g(N_{Z_c})=0.9
\end{equation}
to be $N^{UL}_{Z_c}=15.1$ events.

For the absolute normalization of the $Z_c^{\pm}(3900)$ production rate we
estimated for the same data sample the number of exclusively produced $\Jpsi$
mesons from incoherent exclusive production in
\begin{equation}
\label{reaction2}
\gamma~N \rightarrow \Jpsi~N,
\end{equation}
the cross section of which is known for our range of $\sqrt{s_{\gamma N}}$
\cite{NA14}. The same selection criteria are applied for the exclusive
production of the $\Jpsi$ mesons
\begin{equation}
\label{reaction3}
\mu^+~N \rightarrow \mu^+ \Jpsi~N,
\end{equation}
and $Z_c^{\pm}(3900)$ hadrons.  To separate $\Jpsi$ production and non-resonant
production of dimuons, the dimuon mass spectrum is fitted by a function
consisting of three Gaussian (two to describe the $\Jpsi$ peak and one for the
$\psi(2S)$ peak) and an exponential background under the peaks (see
Fig. \ref{fig:kin}a).  Finally $18.2\times 10^3$ events of exclusive $\Jpsi$
production remain in the sample. The distribution of the squared transverse
momentum $p^2_{T}$ of the $\Jpsi$ (Fig.~\ref{fig:kin3}b) for the exclusive
sample is fitted by a sum of two exponential functions in order to separate the
contributions from exclusive coherent production on the target nuclei and
exclusive production on (quasi-)free target nucleons.  The contribution from
coherent production is found to be 30.3\% for the $^6$LiD target and 38.9\% for
NH$_3$ target (36.1\% averaged over the sample).  The amount of nonexclusive
events in the exclusive incoherent sample is estimated to be about $30\pm10\%$.
Since only the charged pion distinguishes the final state of process in
Eq.~(\ref{reaction}) from the final state of the process in
Eq.~(\ref{reaction2}), the ratio $R_a$ of their acceptances is in a first
approximation equal to the acceptance for this pion. Based on previous COMPASS
measurements and Monte Carlo simulations this ratio is about $R_a=0.5$, averaged
over all setup and target configurations. Thus we obtain the result
\begin{equation}
\label{result}
\frac{BR(Z_c^{\pm}(3900)\rightarrow \Jpsi \pi^{\pm} )\times \sigma_{ \gamma~N
    \rightarrow Z_c^{\pm}(3900)~ N}}{\sigma_{ \gamma~N \rightarrow \Jpsi~ N}
}\Big{|}_{\langle\sqrt{s_{\gamma N}}\rangle=13.8~\GeV} < 3.7\times10^{-3},
\end{equation}
where $BR$ denotes the branching ratio for the $Z_c^{\pm}(3900)\rightarrow \Jpsi
\pi^{\pm}$ decay channel.  Assuming $\sigma_{ \gamma~N \rightarrow \Jpsi~
  N}=14.0\pm1.6_{stat}\pm2.5_{syst}$~nb as measured by the NA14 Collaboration
for $\sqrt{s}_{\gamma N}=13.7~\GeV$ \cite{NA14}, the result can be presented as
\begin{equation}
\label{result2}
BR(Z_c^{\pm}(3900)\rightarrow \Jpsi \pi^{\pm} )\times \sigma_{ \gamma~N
  \rightarrow Z_c^{\pm}(3900)~ N}\Big{|}_{\langle\sqrt{s_{\gamma
      N}}\rangle=13.8~\GeV} < 52~\mathrm{pb}.
\end{equation}

The upper limits for the ratio of the cross sections in intervals of
$\sqrt{s_{\gamma N}}$ are presented in Table~\ref{Zcfit}.
 
The result shown in Eq.~(\ref{result2}) can be converted into an upper limit for
the partial width $\Gamma_{\Jpsi\pi}$ of the decay in Eq.~(\ref{decay}) based on
the VMD model. According to Ref.~\cite{MAIN} the cross section for the reaction
in Eq.~(\ref{reaction}), averaged over the measured $\sqrt{s_{\gamma N}}$
distribution for $\Jpsi\pi^{\pm}$ events is about $\Gamma_{\Jpsi\pi}\times
430$~pb/MeV for $\Lambda_{\pi}=0.6~\GeV$, a free parameter of the $\pi NN$
vertex, yielding

\begin{equation}
\frac{\Gamma_{J/\psi\pi}}{\Gamma_{tot}}\times \sigma_{ \gamma~N \rightarrow
  Z_c^{\pm}(3900)~ N} = \frac{\Gamma_{J/\psi\pi}^2 \times
  430~\mathrm{pb/MeV}}{\Gamma_{tot}}<52~\mathrm{pb}.
\end{equation}
Assuming $\Gamma_{\mbox{tot}}=46$~MeV/$c^2$, we obtain an upper limit
$\Gamma_{J/\psi\pi}<2.4$ MeV/c$^2$.

We estimate the systematic uncertainty of the result in Eq. (\ref{result}) to be
about 30\%, where the main contributions come from limited knowledge of the
acceptance ratio $a$, since the energy spectrum of the expected
$Z_c^{\pm}(3900)$ events is unknown, from systematic effects in estimation of
nonexclusive contamination in the reference $\Jpsi$ sample and from the
background description in the signal range of the $\Jpsi\pi$ spectrum. The
uncertainty of $\sigma_{ \gamma~N \rightarrow \Jpsi~ N}$ measurement by NA14
contributes to Eq. (\ref{result2}), so the total systematic uncertainty of this
result is about 35\%.

\begin{figure}
 \begin{center}
  \includegraphics[width=220px]{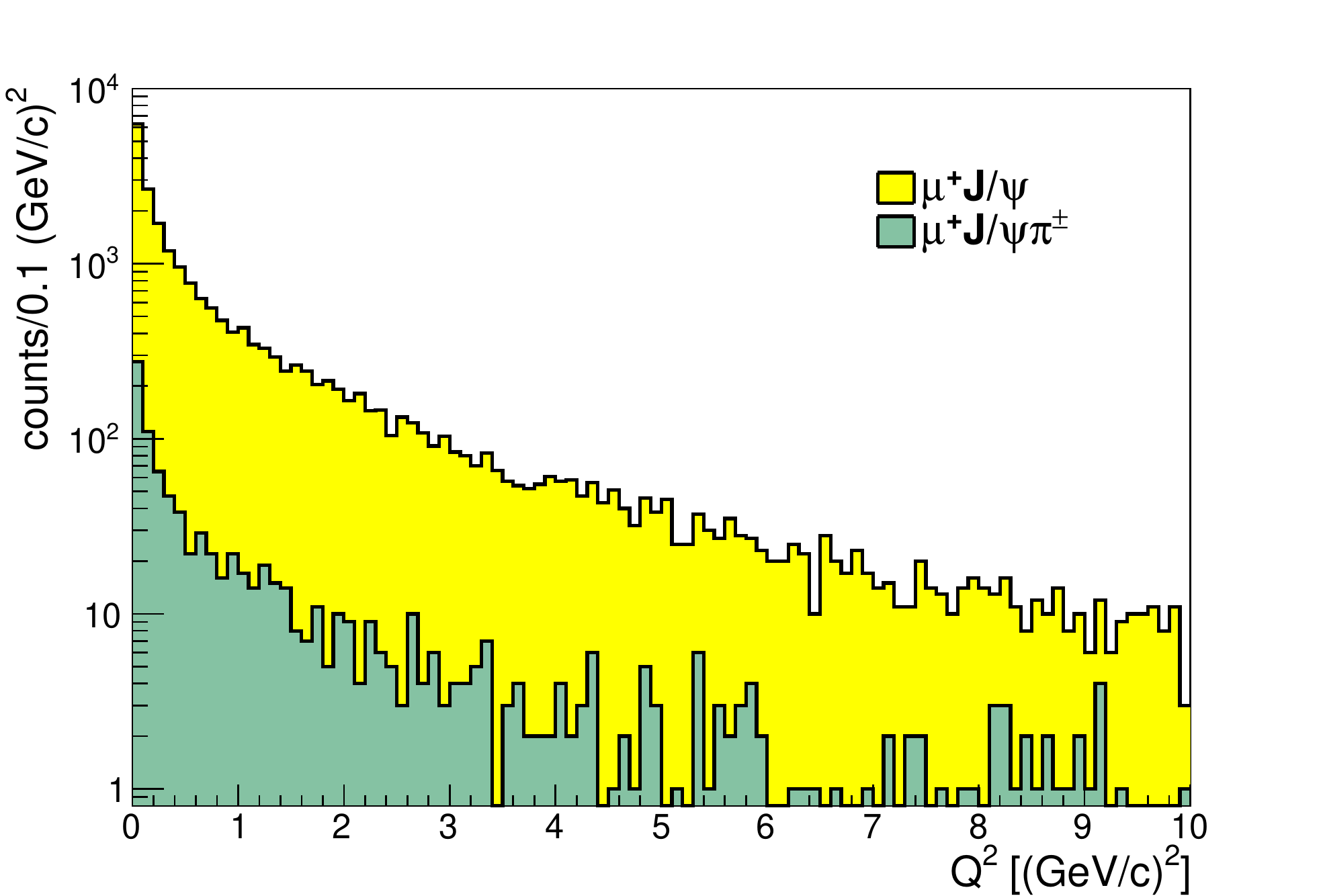}
   \includegraphics[width=220px]{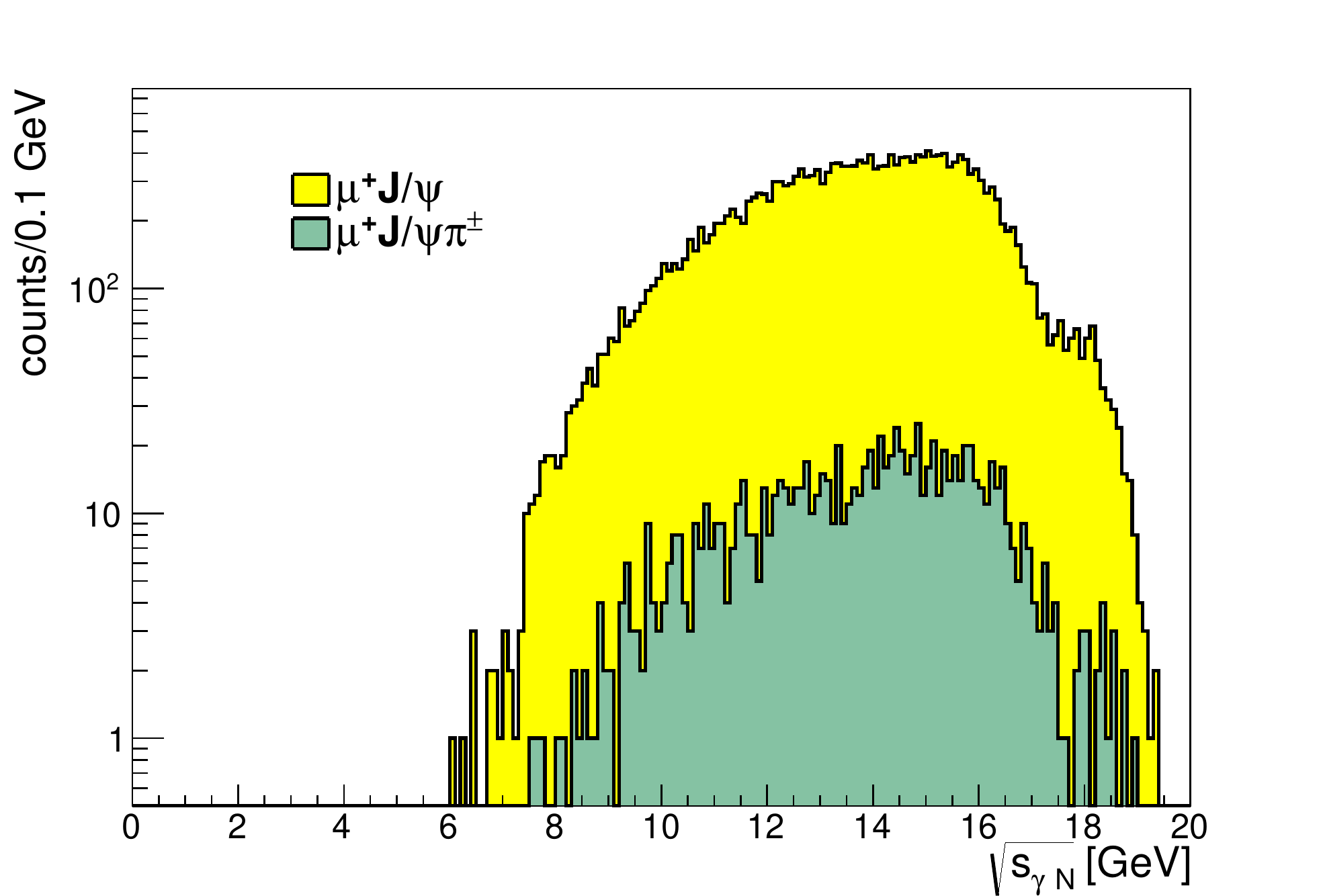}         \\
    (a)\hspace{0.4\textwidth}(b)$\quad$    
 \end{center}
  \caption{\label{fig:kin2} Kinematic distributions for the reactions Eq.~(7)
    (yellow, upper curves) and Eq.~(3) (green, lower curves) (a) $Q^2$, (b)
    $\sqrt{s_{\gamma N}}$.  }
\end{figure}

\begin{figure}
 \begin{center}
   \includegraphics[width=220px]{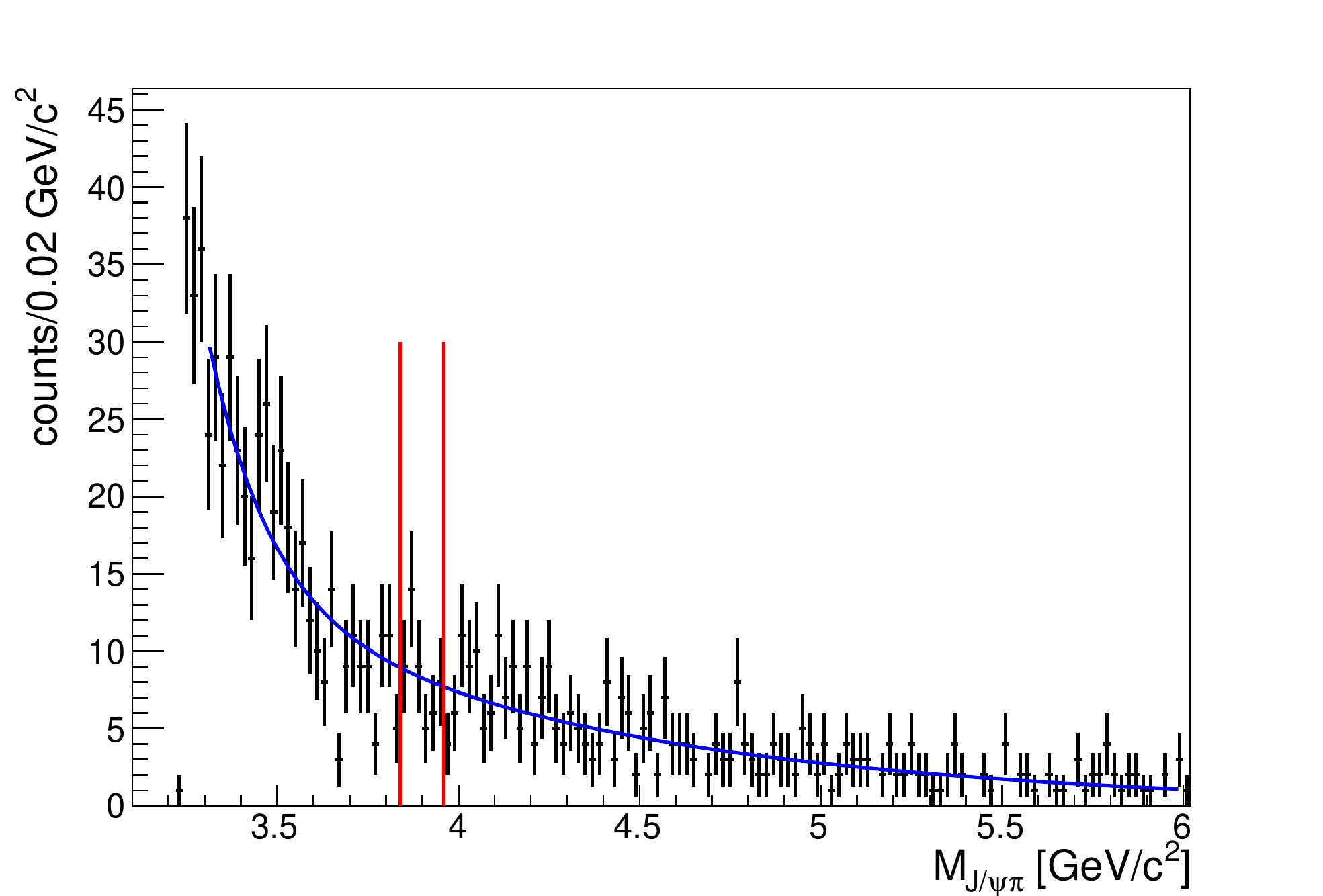}
   \includegraphics[width=220px]{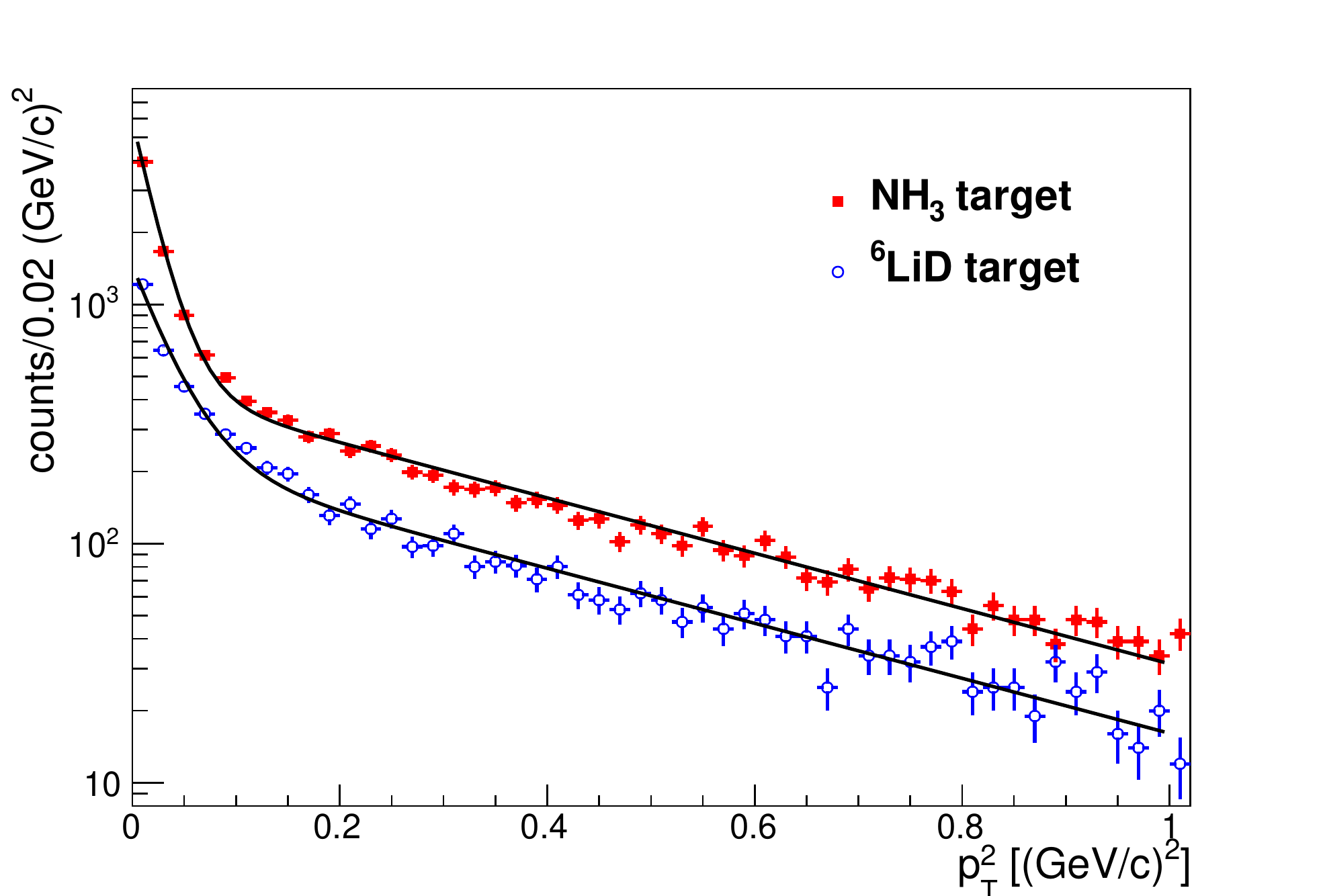}\\ (a)\hspace{0.4\textwidth}(b)$\quad$
 \end{center}
  \caption{\label{fig:kin3} (a) Mass spectrum of the $\Jpsi \pi^{\pm}$
    state. The fitted function is shown as a line. (b) $p^2_T$ distributions for
    exclusively produced $\Jpsi$ mesons off the $^6$LiD (blue, lower) and NH$_3$
    (red, upper) targets.  }
\end{figure}

\begin{table}[htdp]
\caption{Upper limits for $Z^{\pm}_c(3900)$ production rate for intervals of
  $\sqrt{s_{\gamma N}}$.}
\begin{center}
\begin{tabular}{|r@{}l|c|c|c|}
\hline \multicolumn{2}{|c|}{Interval} & $\langle\sqrt{ s_{\gamma N}}~\rangle$,
GeV & $BR(\Jpsi \pi)\times \sigma_{Z_{c}}/\sigma_{\Jpsi\strut}$,
10$^{\strut-3}$\\ \hline \multicolumn{2}{|c|}{Full} & 13.8 & 3.7\\ \hline
&$\sqrt{s_{\gamma N}}<12.3~\GeV$ & 10.8 & 10 \\ $ 12.3~\GeV<$\,&$\sqrt{s_{\gamma
    N}}<14.1~\GeV$ & 13.2 & 3.7 \\ $14.1~\GeV<$\,&$\sqrt{s_{\gamma
    N}}<15.4~\GeV$ & 14.7 & 4.5\\ $15.4~\GeV<$\,&$\sqrt{s_{\gamma N}}$ & 16.4
&6.0\\ \hline
\end{tabular}
\end{center}
\label{Zcfit}
\end{table}

 No signal of exclusive photoproduction of the $Z_c^{\pm}(3900)$ state and its
 decay into $\Jpsi \pi^{\pm}$ was found.  Therefore an upper limit was
 determined for the product of the cross section of this process and the
 relative $Z_c^{\pm}(3900)\rightarrow \Jpsi \pi^{\pm}$ decay probability
 normalized to the cross section of incoherent exclusive photoproduction of
 $\Jpsi$ mesons.  In case $Z_c^{\pm}(3900)$ is a real hadron state, the decay
 channel $Z_c^{\pm}(3900)\rightarrow \Jpsi \pi^{\pm}$ can not be the dominant
 one.  This result is a significant input to clarify the nature of the
 $Z_c^{\pm}(3900)$ state.

We gratefully acknowledge the support of the CERN management and staff as well
as the skills and efforts of the technicians of the collaborating institutions.

\end{document}